\documentclass[letterpaper,english,reprint,aps,prl,superscriptaddress,showpacs,longbibliography]{revtex4-1}

\usepackage[latin9]{inputenc}
\usepackage{amsmath}
\usepackage{amssymb}
\usepackage{wasysym}
\usepackage{graphicx}
\usepackage{color,soul}
\usepackage{physics}
\usepackage{siunitx}
\usepackage{dsfont}
\usepackage{float}
\usepackage[english]{babel}
\usepackage{blindtext}
\usepackage[english,nomargin,inline,marginclue,draft]{fixme}
\pdfpageheight\paperheight
\pdfpagewidth\paperwidth

\usepackage[colorlinks,linkcolor=blue,anchorcolor=blue,citecolor=blue,urlcolor=blue]{hyperref}
\fxusetheme{colorsig}
\FXRegisterAuthor{cg}{acg}{CG}  
\FXRegisterAuthor{th}{ath}{\color{blue}TH}  
\FXRegisterAuthor{ib}{aib}{\color{red}IB} 
\FXRegisterAuthor{sh}{ash}{\color{cyan}SH} 
\FXRegisterAuthor{db}{adb}{\color{green}DB} 
\FXRegisterAuthor{ps}{aps}{PS}


\long\def\symbolfootnote[#1]#2{\begingroup%
\def\thefootnote{\fnsymbol{footnote}}\footnotetext[#1]{#2}\endgroup}

\def\nobreakbefore{%
  \relax\ifvmode\else
    \ifhmode
      \ifdim\lastskip > 0pt\relax
        \unskip\nobreakspace
      \else 
        \nobreakspace
      \fi
    \fi
  \fi
}
\let\oldcite\cite
\renewcommand\cite{\nobreakbefore\oldcite}

\begin{document}
\title{Folded multistability and hidden critical point in microwave-driven Rydberg atoms}

\author{Yu Ma$^{1,2,\textcolor{blue}{\star}}$}
\author{Bang Liu$^{1,2,\textcolor{blue}{\star}}$}
\author{Li-Hua Zhang$^{1,2,\textcolor{blue}{\star}}$}
\author{Ya-Jun Wang$^{1,2}$}
\author{Zheng-Yuan Zhang$^{1,2}$}
\author{Shi-Yao Shao$^{1,2}$}
\author{Qing Li$^{1,2}$}
\author{Han-Chao Chen$^{1,2}$}
\author{Jun Zhang$^{1,2}$}
\author{Tian-Yu Han$^{1,2}$}
\author{Qi-Feng Wang$^{1,2}$}
\author{Jia-Dou Nan$^{1,2}$}
\author{Yi-Ming Yin$^{1,2}$}
\author{Dong-Yang Zhu$^{1,2}$}
\author{Guang-Can Guo$^{1,2}$}
\author{Dong-Sheng Ding$^{1,2,\textcolor{blue}{\dag}}$}
\author{Bao-Sen Shi$^{1,2}$}

\affiliation{$^1$Key Laboratory of Quantum Information, University of Science and Technology of China, Hefei, Anhui 230026, China.}
\affiliation{$^2$Synergetic Innovation Center of Quantum Information and Quantum Physics, University of Science and Technology of China, Hefei, Anhui 230026, China.}

\date{\today}

\symbolfootnote[1]{Y.M, B.L, and L.H.Z contribute equally to this work.}

\symbolfootnote[2]{dds@ustc.edu.cn}

\begin{abstract}
The interactions between Rydberg atoms and microwave fields provide a valuable framework for studying the complex dynamics out of equilibrium, exotic phases, and critical phenomena in many-body physics. This unique interplay allows us to explore various regimes of nonlinearity and phase transitions. Here, we observe a phase transition from the state in the regime of bistability to that in multistability in strongly interacting Rydberg atoms by varying the microwave field intensity, accompanying with the breaking of \( \mathbb{Z}_{3} \)-symmetry. During the phase transition, the system experiences a hidden critical point, in which the multistable states are difficult to be identified. Through changing the initial state of system, we can identify a hidden multistable state and reveal a hidden trajectory of phase transition, allowing us to track to a hidden critical point. In addition, we observe multiple phase transitions in spectra, suggesting higher-order symmetry breaking. The reported results shed light on manipulating multistability in dissipative Rydberg atoms systems and hold promise in the applications of non-equilibrium many-body physics.
\end{abstract}

\maketitle

\section{Introduction}
Due to exaggerated properties of Rydberg atoms \cite{saffman2010quantum,firstenberg2016nonlinear,adams2019rydberg,browaeys2020many}, the Rydberg atoms systems have been shown to exhibit many intriguing characteristics of complex systems, including non-equilibrium phase transitions \cite{lee2012collective,carr2013nonequilibrium,malossi2014full,lesanovsky2014out}, self-organization \cite{helmrich2020signatures,ding2019Phase,klocke2020hydrodynamic}, non-ergodicity and time crystals \cite{gambetta2019,ding2023ergodicity,Wadenpfuhl2023Synchronization, wu2023,liu2024higher,liu2024bifurcation,liu2024microwave}, and non-Hermitian many-body physics \cite{zhang2024exceptional}. Because of interaction-induced nonlinearity, the system settles into one of two stable configurations, depending on the external influences such as electric fields or the density of Rydberg atoms \citep{wade2018terahertz,ding2022enhanced,liu2024microwave,zhang2024early}, resulting in complex bistability \cite{carr2013nonequilibrium,ding2019Phase} or hysteresis loop \cite{zhang2024exceptional}. The stationary states within the bistable regime can be effectively described using a double-well potential \citep{marcuzzi2014universal}. This framework encapsulates the transitions between a high-density phase and a low-density phase. The emergence of bistability in Rydberg atoms has become an intriguing resource to study non-equilibrium many-body physics \citep{carr2013nonequilibrium,malossi2014full,vsibalic2016driven,wade2018terahertz,weller2019interplay}. In contrast to other systems there is no optical cavity feedback \citep{gibbs1976differential,wang2001bistability,wang2001enhanced}; or requirement for cryogenic temperatures \citep{hehlen1994cooperative}. 

It is worth noting that the multiple phases could fold together when Rydberg atoms are driven by microwave fields, thus the system cannot be broken down to just two global states. Within a hysteresis loop, there can be multiple states folded on top of each other, resulting from high-order nonlinearity and high-order symmetry breaking. This phenomenon is often observed in complex systems characterized by non-linear dynamics, feedback loops, and multiple interconnected variables. As parameters are varied within a certain range, different states emerge and coexist, resulting in complex hysteresis loops and multistability across various variables in different systems, including biological systems, chemical reactions, ecosystems and climate dynamics \cite{may1977thresholds,simoyi1982one,paillard1998timing,laurent1999multistability,pisarchik2014control}. This multistable effect has been predicted and investigated in coupled atom-cavity systems \citep{kitano1981optical,cecchi1982observation,joshi2003optical,sheng2012realization} and a semiconductor microcavity \citep{gippius2007polarization,paraiso2010multistability,cerna2013ultrafast,goblot2019nonlinear}. The mutually fold in a multistable system could exhibit bistability within the hysteresis loop, which results in hidden multistability where the number of output states is less than that of the steady states of system. This makes it challenging to identify all possible states in systems that contain hidden multistability, making it a compelling topic for exploration and theoretical investigation \cite{bi2024folding}.

\begin{figure*}
\centering
\includegraphics[width=1\linewidth]{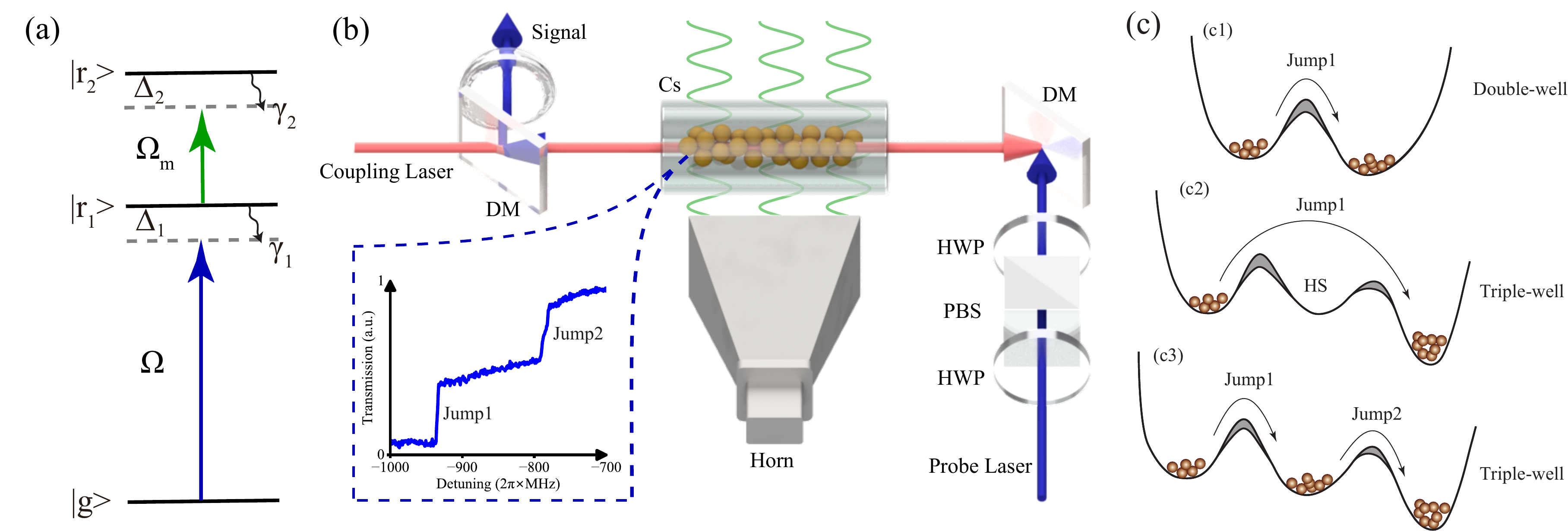}
\caption{\textbf{Schematic of multistability in Rydberg atoms}. (a) Energy level diagrams. A laser field excites the atoms from ground state $\ket{g}$ to the Rydberg state $\ket{r_1}$ with Rabi frequency $\Omega$ and detuning $\Delta_{1}$. Another microwave field couples the Rydberg states $\ket{r_1}$ and $\ket{r_2}$ with Rabi frequency $\Omega_m$ and detuning $\Delta_{2}$.  $\gamma_{1}$ and $\gamma_2$ are decay rates of the Rydberg states $\ket{r_1}$ and $\ket{r_2}$. 
(b) Schematic diagram of the experimental setup. The probe beam is incident opposite to the coupling beam and overlap in the centre of Cesium vapor cell. The Cesium vapor cell is shinned by a microwave field from a horn. DM, dichroic mirror; PBS, polarizing beam splitter; HWP, half-wave plate; Cs, Cesium. (c) The phase transition from the state in bistable regime (double-well potential, c1) to the hidden state (HS) in multistability (triple-well potential, c2) and to the state in unhidden multistability (triple-well potential, c3).}
\label{Fig.1}
\end{figure*}

In this work, we observe a phase transition from bistability to multistability in a thermal Cesium Rydberg atoms system under a microwave field driving. At low microwave field intensities, the strong interaction between Rydberg atoms induces bistability. However, when the microwave field intensity is increased, the Rydberg energy levels undergo splitting due to the Autler-Townes effect \cite{sedlacek2012microwave,liu2023electric}. This breaks $\mathbb{Z}_{3}$-symmetry and results in a phase transition from bistability to multistability. This process reveals a hidden critical point, near which a third, folded state is concealed within the bistability hysteresis loop (which only presents two observable states). Consequently, multiple non-equilibrium steady states coexist within the same parameter range. In this context, the system's behavior cannot be reduced to merely two stable states. By carefully tuning the system parameters, we demonstrate that these multiple non-equilibrium steady states can be distinctly identified and distinguished, including hidden states and sublevels-dependent steady states. The reported findings on the phase transition and the folding effect of multistability in the Rydberg atoms system hold significant promise for studying and characterizing intriguing non-equilibrium phases in the realms of nonlinearity and many-body physics.

\section{Physical model}
To demonstrate the phase transition from the state in bistability to that in multistability, we build a simplified physical model to uncover the underlying mechanism of experimental observations. The model consists of $N$ interacting three-level atoms with a ground state $\ket{g}$ and the Rydberg states $\ket{r_1}$ and  $\ket{r_2}$ (with a decay rate of $\gamma_1$ and $\gamma_2$). The atoms are coupled by a laser with a Rabi frequency $\Omega$ and detuning $\Delta_1$. For the experiment, this process can be realized by an off-resonance two-photon Raman excitation scheme, see more details in appendix. With a large detuning $\Delta_p$, the intermediate state $\ket{e}$ is adiabatically eliminated. The Rydberg states  $\ket{r_1}$ and $\ket{r_2}$ are driven by a resonant microwave field with a Rabi frequency $\Omega_m$, a frequency $\omega_m$, and a detuning $\Delta_2$. 

The energy diagram of the model is depicted in Fig.\ref{Fig.1}(a). The experimental setup can be found in Fig.\ref{Fig.1}(b), in which the insert figure is the measured transmission spectrum with two jumps. The two jumps exhibit the characteristic of multistability. To describe the underlying mechanism behind the observations, we plot different potential wells to explain it. Here, one jump corresponds to the transition between two stationary states in a double-well potential \citep{marcuzzi2014universal}, while two jumps reveal the transitions in a triple-well potential, as illustrated in Fig.\ref{Fig.1}(c). The interaction between Rydberg atoms is governed by the van der Waals interaction strength $V = {C_6}/{r^6} $ [where $C_6$ is coefficient and $r$ represents the distance between the Rydberg atoms]. The Hamiltonian of system is:
\begin{align*}
    \hat{H} & =\frac{1}{2}\sum_{i}\left(\Omega\sigma_{i}^{gr_1}+\Omega_{m}\sigma_{i}^{r_1r_2}+h.c.\right)\\ & +\frac{1}{2}\sum^{m,n=1,2}_{i\neq j}V_{ij}^{r_mr_n}n_{i}^{r_m}n_{j}^{r_n}+\sum_{i}\left(\Delta_{1}n_{i}^{r_1}+\Delta_{2}n_{i}^{r_2}\right)
\end{align*}
where $\sigma_{i}^{gr_1}$ ($\sigma_{i}^{r_1r_2}$) are the $i$-th atom transition between $\left| g \right\rangle$ ($\left|  r_1 \right\rangle$) and $\left|  r_1 \right\rangle$ ($\left|  r_2 \right\rangle$) ; $n_{i}^{r_1}$ and $n_{i}^{r_2}$ are the population operators of states $\left|  r_1 \right\rangle$ and $\left|  r_2 \right\rangle$; $V^{r_m,r_n}_{ij}$ represents the interactions between Rydberg atoms. 

We calculate the steady-state solution of the Lindblad master equation \(\dot{\rho} = -i[\hat{H},\rho] + \mathcal{L}[\rho]\) under the condition \(\dot{\rho} = 0\), employing the mean-field approximation \(\Delta_1 \rightarrow \Delta_1 + \Delta_{\text{shift}} - V\rho_{r_1r_1}\), see more information in appendix. Here, $\mathcal{L}[\rho]$ is the Lindblad term, \(\Delta_{\text{shift}}\) represents the microwave fields-dependent energy shift, which is proportional to the Rabi frequency of the microwave field. For simplicity, we set \(\Delta_{\text{shift}} = \lambda \Omega_m\) with \(\lambda = 1.8\) as an ionization-induced shift coefficient. Consequently, we obtain a series of real roots for the Rydberg population \(\rho_{r_1r_1}\) as a function of \(\Delta_1\).

\begin{figure*}
\centering
\includegraphics[width=1.04\linewidth]{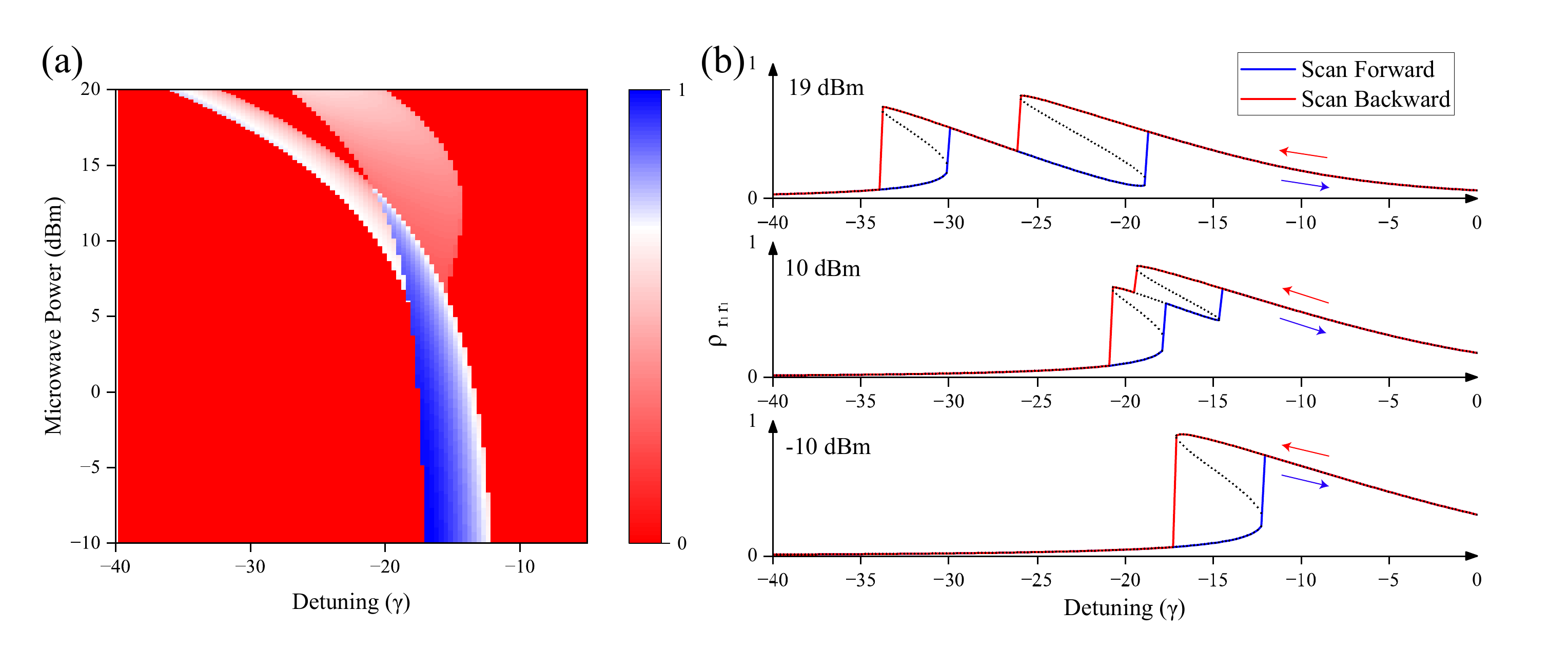}
\caption{\textbf{Theoretical phase diagram and multistability}. (a) Theoretical phase diagram depicting the variation of $\rho_{r_1r_1}$ with respect to microwave field power and laser detuning under forward and backward scanning detuning. The colored boundary lines on the left side of the figure correspond to the boundaries of the phase diagram. As the microwave field power gradually increases, the population difference $\rho_{r_1r_1}$ in the bistability regime becomes split, eventually leading to the emergence of multistability. The color bar is to show the mapping of values to colors in the displayed data. (b) The spectrum of \(\rho_{r_1r_1}\) is presented for varying microwave field intensities with dotted lines. The blue and red lines show the behaviors of forward and backward scanning detuning, respectively, while the dotted lines that are not covered by solid lines at all represents solutions to unstable states. The lower panel of (b) illustrates the bistability range, while the middle and upper panels showcase the spectrum in a multistable regime. The dashed lines in these figures indicate the calculated solutions obtained from the Lindblad master equation. In these calculations, the parameters are set as $\Gamma_1=8 \gamma$, $\Gamma_2=\gamma$, $\Delta_2=1.5 \gamma$, and $V = -200 \gamma$.}
\label{Fig2}
\end{figure*}

\section{Phase diagram and multistability}
We plot the phase map of Rydberg population $\rho_{r_1r_1}$ and spectrum versus the detuning $\Delta_1$ and the intensity of microwave field $P$, as shown in Fig.~\ref{Fig2}. Figure.~\ref{Fig2}(a) shows the Rydberg population difference $\delta\rho_{r_1r_1}=\rho^B_{r_1r_1}-\rho^F_{r_1r_1}$, where $\rho^B_{r_1r_1}$ and $\rho^F_{r_1r_1}$ are the corresponding populations of Rydberg atoms under forward and backward scanning $\Delta_1$, respectively. By increasing the intensity of microwave field from -10 dBm to 20 dBm, we can find that there is a complex phase diagram characterizing the regime of hysteresis loops. The regime of hysteresis loops within bistability starts to split at $P$= 6 dBm and becomes double bistability when $P>$ 13.5 dBm. Figure.~\ref{Fig2}(b) shows the calculated spectrum under $P$ = -10 dBm (down panel), 10 dBm (middle panel), and 19 dBm (up panel). The combination of the blue and red lines forms complex hysteresis loops. For the regime in bistability, the population $\rho_{r_1r_1}$ has three solutions [for example, $\Delta_1 = -15 \gamma$ in the down panel of Fig.~\ref{Fig2}(b)], while for multistability, there is a case where there are 5 solutions, see the population $\rho_{r_1r_1}$ at $\Delta_1 = -18.5 \gamma$ in the middle panel of Fig.~\ref{Fig2}(b). 

The physical picture of system is captured by a mean-field potential in dissipative Rydberg atoms \citep{marcuzzi2014universal}. When the Rydberg atoms interact with weak microwave fields, the \( \mathbb{Z}_{2} \)-symmetry is broken because of strong interaction between Rydberg atoms, leading to a nonequilibrium phase transition between two distinct stationary states of system in a self-organization configuration \cite{carr2013nonequilibrium,ding2019Phase}. This corresponds to a single jump in the spectrum associated with a transition between each well in the double-well potential, as shown in the upper panel of Fig.~\ref{Fig.1}(c). The multistability indicates the presence of a higher-order phase transition, which involves the breaking of \( \mathbb{Z}_{3} \)-symmetry, allowing the system to settle into one of three distinct states. When scanning \( \Delta_1 \), we observe two jumps in the spectrum, corresponding to two transitions within a triple-well potential, as depicted in the lower panel of Fig.~\ref{Fig.1}(c).

\begin{figure*}
\centering
\includegraphics[width=1.04\linewidth]{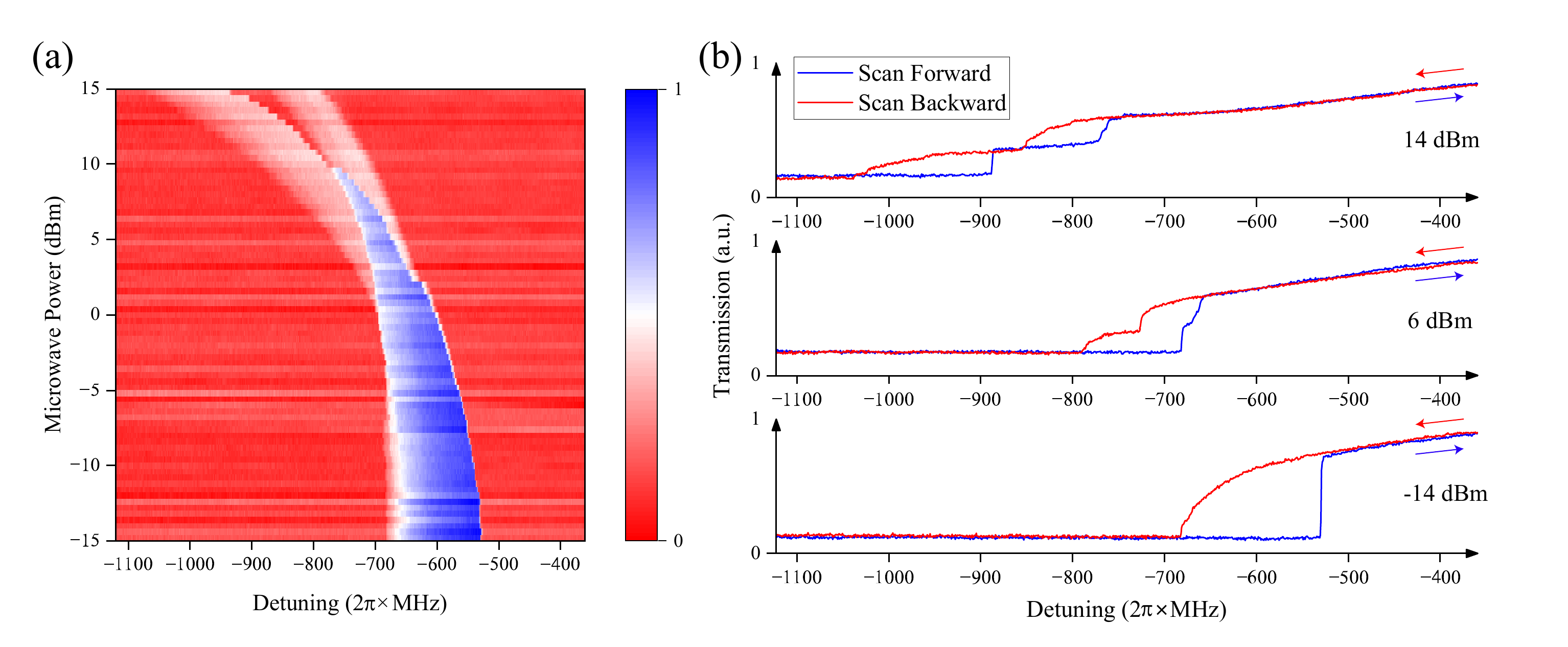}
\caption{\textbf{Measured phase diagram and multistability}. (a) Measured phase diagram of difference of probe transmission versus the microwave field power and laser detuning under forward and backward scanning detuning. Here "Microwave Power" refers to the output microwave power on the signal generator settings . As the microwave power increases, the transmission difference under two scanning directions becomes split gradually from a sharp band. The color bar is to show the mapping of values to colors in the displayed data. (b) The spectrum of probe light with different microwave field power. Blue and red lines correspond to forward and backward scanning detuning. Spectra in (b) are extracted as representation of different transition stages from (a) at different microwave powers. In this process, we set the frequency of the microwave field as $\omega_m=2\pi\times$ 26.17 GHz.}
\label{Fig3}
\end{figure*}

In the experiment, we utilized a two-photon electromagnetically induced transparency (EIT) scheme \cite{harris1990nonlinear,mohapatra2007coherent,petrosyan2011electromagnetically} to excite atoms to a Rydberg state in a cesium thermal vapor, as shown in the experimental configurations in Fig.~\ref{Fig.1}(b). Further information regarding the laser beams and parameters of microwave fields can be found in appendix. The population of Rydberg atoms can be monitored by the response of the atoms to the probe field, allowing for the measurement of the EIT spectrum under various conditions. Bistability and multistability can be achieved by scanning the detuning of coupling field in both forward and backward directions. The regimes of bistability and multistability are determined by subtracting the spectrum from the backward scan and that from the forward scan $\delta T = T^B-T^F$, where $T^B$ and $T^F$ are the corresponding probe transmission under backward and forward scanning $\Delta_1$.

We measure the EIT spectra under varying microwave powers, resulting in a transmission difference phase diagram, as shown in Fig.~\ref{Fig3}(a). In Fig.~\ref{Fig3}(a), as the microwave power $P$ increases from -15 dBm to 15 dBm, the transmission difference $\delta T$ begins to diverge at $P$ = 1 dBm, and completely separates with no overlapping region at $P$ = 10 dBm. This splitting behavior aligns well with the theoretical predictions in Fig.~\ref{Fig2}(a).  In Fig.~\ref{Fig3}(b), we display typical spectra corresponding to different transition stages at the microwave powers of 14 dBm, 6 dBm, and -14 dBm. As the microwave power is scanned, tristability emerges from the bistability spectrum, ultimately evolving towards separable double bistability. 

\begin{figure*}
\centering
\includegraphics[width=1\linewidth]{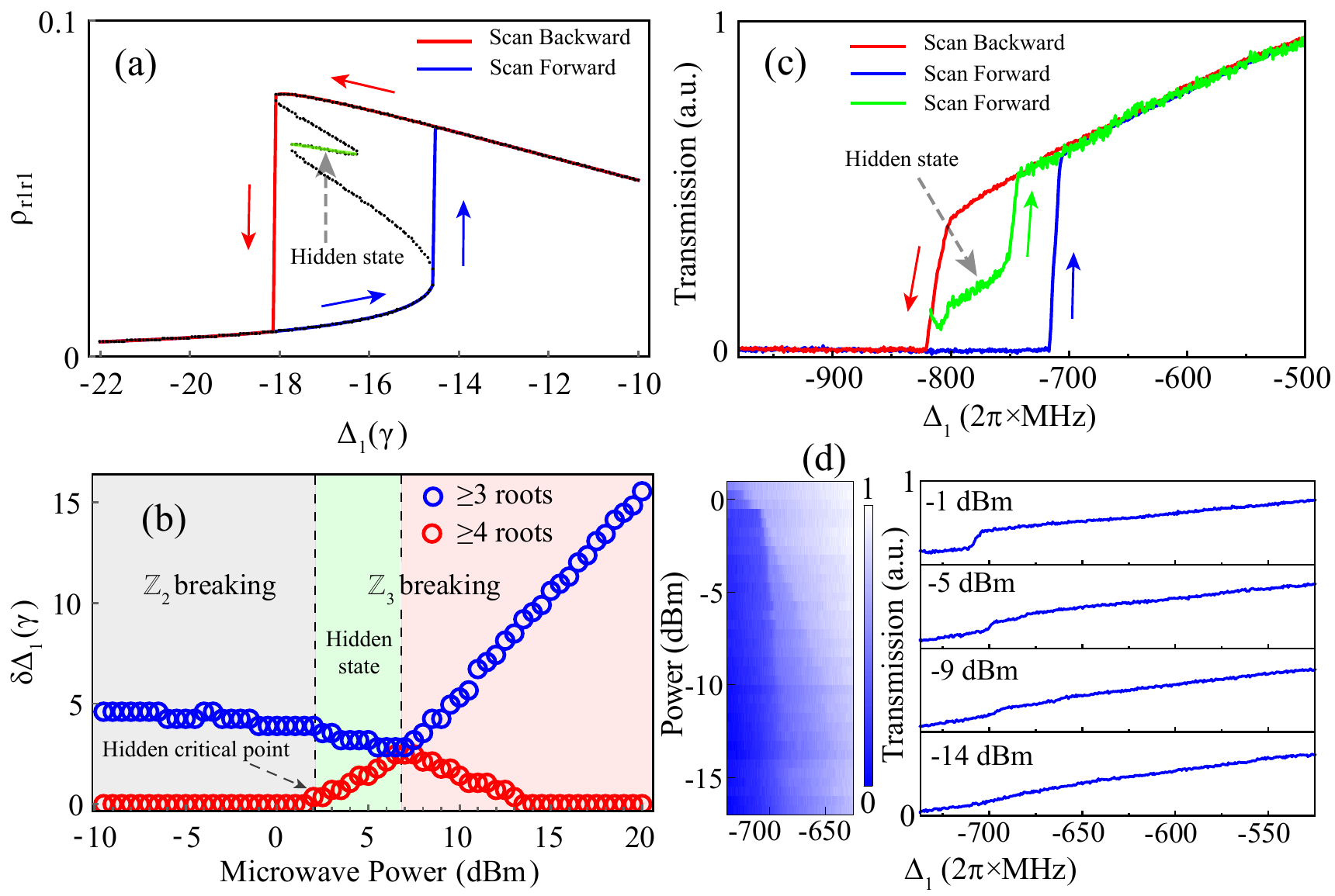}
\caption{\textbf{Hidden critical point and multistability}. (a) Calculated EIT spectrum under forward (blue) and backward (red) scanning $\Delta_1$. The population lines under these two scan cases form a bistable hysteresis loop. Within the bistability regime, there is a hidden multistable state that cannot be obtained by scanning $\Delta_1$. (b) Scatter plot of $\delta\Delta_1$ with $\rho_{r_1r_1}$ having \(\geq 3\) roots (blue circle) and \(\geq 4\) roots (red circle) respectively against microwave power, and here $\delta\Delta_1$ represents the whole span of $\Delta_1$ with certain number of roots. The region of pure bistability (with \( \mathbb{Z}_{2} \)-symmetry breaking) is indicated by the gray area, while the region of multistability (with \( \mathbb{Z}_{3} \)-symmetry breaking) is highlighted in green and red. The concealed multistable state appears as bistability in observations denoted by the green color. (c) Measured hidden multistable state. The hidden state [marked by the shaded arrow] can be identified by scanning the detuning at the starting point $\Delta_1$. The trajectory of probe transmission is shown by the green color, in which there is a hidden jump that corresponds to the phase transition from the hidden state to the unhidden state (with a high transmission level). (d) Measured the phase transitions with varying the microwave power $P = 1$ dBm to  $P = -17$ dBm. The left panel is the phase diagram, and the right panel represents the transmissions under the microwave power $P = -1$ dBm, -5 dBm, -9 dBm, and -14 dBm, respectively. The point where the phase transition disappears corresponds to the hidden critical point. The color bar is to show the mapping of values to colors in the displayed data. Based on these results, we can identify a hidden critical point at which the \( \mathbb{Z}_{3} \)-symmetry begins to break. In this process, we set the frequency of the microwave field as $\omega_m = 2\pi\times$ 26.12 GHz.}
\label{Fig4}
\end{figure*}

\section{Hidden critical point}
Among these phase diagrams, the transition point in spectrum where tristability is on the verge of appearing yet remains concealed within the bistability regime is particularly intriguing. This gives rise to a hidden critical point, wherein the occurrence of phase transition is challenging to detect through a simple scan of \(\Delta_1\). In this context, a hidden critical point refers to a condition in a non-equilibrium system where multiple stable states coexist, but the transitions between these states remain inaccessible. 

To analyze this effect, we plot the calculated EIT spectrum at $P=4$ dBm, as shown in Fig.~\ref{Fig4}(a). The population $\rho_{r_1r_1}$ exhibits bistability under forward (blue) and backward (red) scanning of \(\Delta_1\), forming a hysteresis loop indicative of bistability in the population lines. This bistability suggests that within a certain parameter range, the system can reside in one of two stable states depending on the scanning direction. However, within the bistable regime, there exists a region at which the system hidden multistable states are not directly accessible by scanning \(\Delta_1\) in forward and backward directions [as represented by the green line in Fig.~\ref{Fig4}(a)], indicating the presence of more complex underlying dynamics.

To detect the presence of the hidden state within the multistability regime, we extract the maximal detuning difference \(\delta\Delta_1\) under the conditions of $\rho_{r_1r_1}$ having \(\geq 3\) roots and \(\geq 4\) roots. \(\delta\Delta_1\) is regarded as an order parameter, the onset of the continuous phase transition is characterized by a broken \( \mathbb{Z}_{3} \)-symmetry accompanied by a nonzero \(\delta\Delta_1\). We then plot \(\delta\Delta_1\) against the microwave power \(P\), as shown in Fig.~\ref{Fig4}(b). The phase diagram in Fig.~\ref{Fig4}(b) highlights regions where the system exhibits multiple roots: specifically, \(\geq 3\) roots (blue circles) and \(\geq 4\) roots (red circles). The phase diagram is divided into distinct regions: the gray area represents pure bistability, characterized by \( \mathbb{Z}_{2} \)-symmetry breaking, while the green and red areas indicate regions of multistability, where \( \mathbb{Z}_{3} \)-symmetry breaking occurs. 

Interestingly, the concealed multistable state, associated with the \( \mathbb{Z}_{3} \)-symmetry, manifests as hysteresis loop of bistability in observations both in theory and experiment. This is denoted by the green region in the phase diagram, as shown in Fig.~\ref{Fig4}(b). The hidden critical point, where the transition from \( \mathbb{Z}_{2} \) to \( \mathbb{Z}_{3} \)-symmetry breaking occurs, plays a crucial role in understanding the full dynamics of the system's behavior. How to identify this hidden critical point provides deeper insights into the high-order symmetry-breaking mechanisms and the emergence of multistability in such nonlinear systems.

In the experiment, it is challenging to identify the hidden state enclosed by the hysteresis loop, as analyzed above. The hysteresis loop forms because the state of atoms is influenced not only by their current state but also by their past states [see the red and blue lines in Fig.~\ref{Fig4}(c)]. Thus, by adjusting the current state of system, we can disrupt the enclosed loops. We modified the initial conditions by scanning the detuning at a starting point \(\Delta_1 = -2\pi \times 817\) MHz and measured the EIT spectrum. Here, the starting point is set near the boundary of the loop, thus the system is inclined to reach the nearest steady state point (the hidden state). The trajectory of this case can be seen by the green line in Fig.~\ref{Fig4}(c), in which the dynamics of the hidden state can be mapped fully. 

\begin{figure*}
\centering
\includegraphics[width=1.04\linewidth]{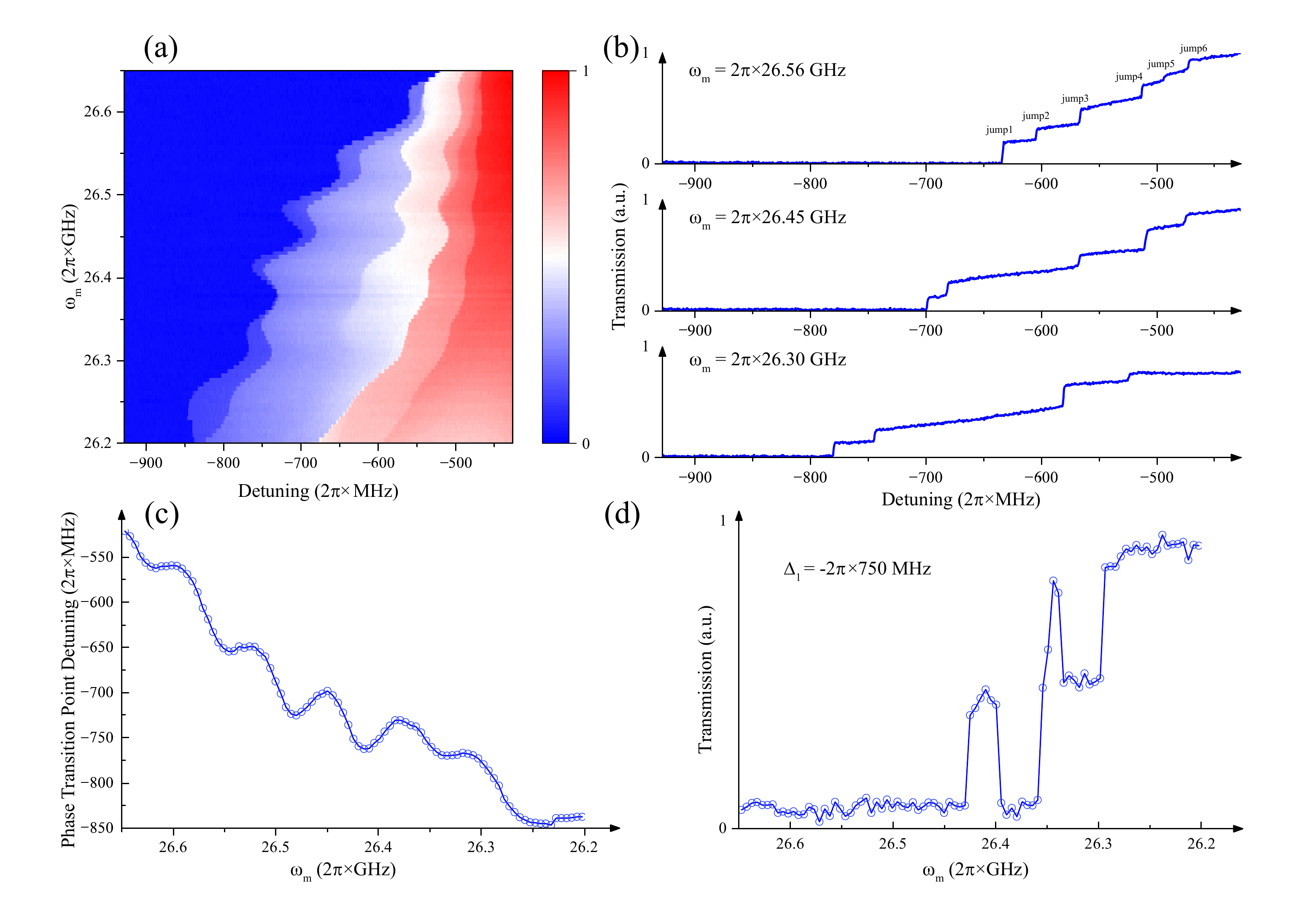}
\caption{\textbf{Multiple jumps in spectrum}. (a) The measured EIT spectrum as a function of microwave frequency $\omega_m$ [from $\omega_m/2\pi=26.2$ GHz to $\omega_m/2\pi=26.65$ GHz] displays distinct transmission stages, with varying color contrasts that indicate multiple phase transitions within the spectra. In this case, we scan $\Delta_1$ with a forward direction. (b) The measured EIT spectra for microwave frequencies of \(\omega_m/2\pi = 26.56\) GHz, \(\omega_m/2\pi = 26.45\) GHz, and \(\omega_m/2\pi = 26.30\) GHz, respectively, reveal multiple jumps in the spectra. There are seven, five, and four jumps in up, middle, and down panel spectrum, respectively. (c) The oscillated effect on first phase transition point. (d) The measured transmission versus microwave frequency \(\omega_m\) while fixing \(\Delta_1 = -2\pi \times 750\) MHz. }
\label{Fig5}
\end{figure*}

By this way, we measure the EIT spectrum versus the microwave power $P$ and obtain the phase diagram [Fig.~\ref{Fig4}(d)], in which there are typical stage-features of single and double hidden phase transitions. These correspond to the transitions from hidden stationary state to a normal stationary state. The right panel of Fig.~\ref{Fig4}(d) shows the spectra of the gradual disappeared phase transitions under different microwave powers. The decrease of the microwave power reduces the gap between the hidden state and the state at high transmission level, leading to the decreased slope at the phase transition point. Through analyzing the data in Fig.~\ref{Fig4}(d), we obtain the position of the hidden critical point $P_c=-14\pm0.5$ dBm.

\section{Multiple jumps in spectrum}
To investigate the scenario of higher-order symmetry breaking, we measure the transmission of probe field as a function of the microwave frequency \( \omega_m \), resulting in a phase diagram depicted in Fig.~\ref{Fig5}(a). In this figure, multiple phase transitions are evident, as indicated by distinct color variations. Figure~\ref{Fig5}(b) presents the examples of transmission spectra for microwave frequency \(\omega_m/2\pi = 26.56\) GHz, \(\omega_m/2\pi = 26.45\) GHz, and \(\omega_m/2\pi = 26.30\) GHz. The results reveal a clear presence of four, five, and six jumps in the EIT spectra, respectively.

The presence of microwave fields introduces distinct frequency shifts in the sublevels of $m_j$ due to Autler-Townes effect, allowing for the observation of multiple transitions or jumps in the EIT spectrum as we adjust the microwave detuning $\Delta_2$. In system, the energy levels associated with different $m_j$ states shift due to the distinct Autler-Townes effect, leading to a situation where the population of these Rydberg sublevels can differ significantly depending on the microwave detuning. As $\omega_m$ is varied, the microwave field induces resonant interactions with these shifted energy levels, resulting in observed multiple jumps in the EIT spectrum, see the jump1 to jump6 in the up panel of Fig~\ref{Fig5}(b). Each jump corresponds to a specific transition between a sublevel influenced by the microwave field. The discrete feature of these transitions reflects the fine structure of the energy levels in the presence of the external field. 

All phase transition points in the phase diagram exhibit oscillatory effects versus the microwave frequency \(\omega_m\). Figure~\ref{Fig5}(c) illustrates the dynamics of the first phase transition point, showcasing a pronounced oscillation effect. The variations in \(\omega_m\) induce resonance and off-resonance conditions between the microwave field and the Rydberg sublevels of \(m_j\), resulting in distinct population distributions. This leads to jumps at different detunings, culminating in the observed oscillatory phenomenon. In addition, we present a transmission spectrum plotted as a function of \(\omega_m\) at a detuning of \(\Delta_1 = -2\pi\times750\) MHz, as shown in Fig~\ref{Fig5}(d). Here, several counterintuitive phase transition points can be observed, where the transmission exhibits abrupt increases and decreases, resulting in all-microwave multistate switching \cite{sheng2012realization}. This behavior arises from the varying coupling strengths under both resonance and off-resonance conditions, as discussed above. 

The observation of $k$ distinct jumps in the EIT-spectrum indicates the breaking of \( \mathbb{Z}_{k+1} \)-symmetry. For example, in Fig~\ref{Fig5}(b), there are six jumps in spectrum that corresponds to \( \mathbb{Z}_{7} \)-symmetry breaking. This symmetry pertains to the cyclic feedback due to the interaction-induced nonlinearity, where the energy level structure allows for $k+1$ equivalent states or potential wells. As $\omega_m$ is varied, the selective coupling of these states leads to high-order symmetry breaking, where certain paths of state transitions become energetically favorable while others do not. This breaking of symmetry manifests as the observed jumps, each corresponding to a different processes of population transfer among the $k+1$ accessible sublevels. The experimental observations of such symmetry breaking are promising, as they open avenues for exploring this similar symmetry breaking in other high-dimensional systems. 

\section{Discussions}
Studying microwave-driven multistability and their transitions in Rydberg atom systems offers significant insights in other aspects induced bistability or multistability both theoretically and experimentally \citep{lee2012collective, marcuzzi2014universal, weimer2015variational, vsibalic2016driven,levi2016quantum,ding2021epidemic}. For instance, random particle motion may also drive non-equilibrium phase transitions \citep{vsibalic2016driven}. In addition, quantum fluctuations can facilitate transitions between multiple stable collective states, as suggested by classical mean-field theory \citep{lee2012collective} and the variational principle \citep{weimer2015variational}. In these methods and technologies, the effective coupling between Rydberg atoms and microwave fields enables more precise control over phase transitions and multistability.

Furthermore, exploring hidden critical points in Rydberg atom systems could illuminate broader questions within driven-dissipative many-body systems, particularly in the context of high-order symmetry breaking, including applications of identifying some exotic hidden phases \cite{bi2024folding}. As a quantum many-body system may have multiple possible hidden critical points, it is promising to study the mechanism of these hidden critical points \cite{haldar2023hidden}. The developed methods to find the hidden critical points in this work allow us to uncover and understand previously unknown phases of matter and unlocking new frontiers in quantum many-body physics.

In summary, we investigate a phase transition from the state in the regime of bistability to multistability under varying microwave field intensities in a thermal Cesium Rydberg atoms system. By tuning the coupling between Rydberg atoms and the Autler-Townes effect, we uncover a rich hysteresis trajectories of multistability, characterized by distinct identifiable stationary states, and non-trivial hidden states. This research underscores that the system cannot simply be confined to two observable states, as multiple stationary states coexist, enriching the understanding of phase transitions in driven-dissipative many-body systems. In addition, the emergence of a hidden critical point and the identification of a hidden state within the bistability hysteresis loop highlight the complexity of the phase behavior at play in Rydberg atoms systems. The observation of multiple jumps in the EIT spectrum serves as a clear signature of  \( \mathbb{Z}_{k} \)-symmetry breaking, suggesting exotic phases in Rydberg many-body systems.

\hspace*{\fill}

\section*{Acknowledgements}
We acknowledge funding from the National Key R and D Program of China (Grant No. 2022YFA1404002), the National Natural Science Foundation of China (Grant Nos. U20A20218, 61525504, and 61435011), and the Major Science and Technology Projects in Anhui Province (Grant No. 202203a13010001). Y.M. conducted the physical experiments with B.L. and L.-H.Z.  D.-S.D. and Y.J.W developed the theoretical model. The manuscript was written by D.-S.D. All authors contributed to discussions regarding the results and the analysis contained in the manuscript. D.-S.D. conceived the idea and supported this project. 

\section*{APPENDIX A: Details of the experimental setup}
We employed a two-photon excitation scheme to drive atoms in the ground state to a higher Rydberg state. In this process, a probe field facilitates the transition from the state \(\ket{6S_{1/2}, F=3}\) to the excited state \(\ket{7P_{3/2}, F=4}\), while a coupling field is responsible for driving the transition from \(\ket{7P_{3/2}, F=4}\) to the Rydberg state \(\ket{31D_{5/2}}\) in Cesium atoms. To maximize the interaction strength between Rydberg atoms, we heated the Cesium atom vapor cell to be $117^\circ$. Specifically, a microwave field operating at $2\pi\times$26.17 GHz is applied to couple the transition between the Rydberg state \(\ket{31D_{5/2}}\) and another excited state \(\ket{32P_{3/2}}\), with a detuning of $2\pi\times$331 MHz. This configuration enables precise control over the energy levels and enhances the coherence of the atomic transitions.

As illustrated in Fig.~\ref{Fig.1}(b), we created an excitation region within the vapor cell using a laser beam at a wavelength of 456 nm (Toptica DLpro laser), which serves as the first photon in our two-photon excitation process. Simultaneously, a counter-propagating beam at 1072 nm (Toptica DLpro laser and a fiber amplifier) is utilized to facilitate the second photon transition, ensuring optimal overlap and interaction between the two beams within the vapor cell. The microwave field is emitted from a horn antenna, strategically positioned to be perpendicular to the direction of lasers propagation. After the excitation process, the probe light that passes through the vapor cell is collected as a signal. This collected light can be used to extract the transmission characteristics of the atomic medium, allowing us to analyze the effect of Rydberg excitation on the atomic population and assess the transmittance of the atoms under different experimental conditions. By collecting spectra while scanning the microwave power in both directions of coupling detuning, we obtain the data for constructing Fig.~\ref{Fig3}.

\section*{APPENDIX B: Lindblad master equation}
As our system can be modeled using a three-level atomic configuration, we construct the Lindblad master equation in the absence of interactions, resulting in the following equations:
\begin{align}
\dot{\rho}_{g g} \ \  = &\ i \frac{\Omega}{2}\left(\rho_{r_1 g}-\rho_{g r_1}\right)+\gamma_1 \rho_{r_1 r_1}, \\
\dot{\rho}_{r_1 r_1} = &\ i\frac{\Omega}{2}\left(\rho_{g r_1}-\rho_{r_1 g}\right)+i \frac{\Omega_m}{2}\left(\rho_{r_2 r_1}-\rho_{r_1 r_2}\right)\notag\\&\,-\gamma_1 \rho_{r_1 r_1}+\gamma_2 \rho_{r_2 r_2}, \\
\dot{\rho}_{r_2 r_2} = &\ i \frac{\Omega_m}{2}\left(\rho_{r_1 r_2}-\rho_{r_2 r_1}\right)-\gamma_2 \rho_{r_2 r_2}, \\
\dot{\rho}_{r_2 r_1} = &\ i \frac{\Omega_m}{2}\left(\rho_{r_1 r_1}-\rho_{r_2 r_2}\right)-i \frac{\Omega}{2} \rho_{r_2 g}\notag\\&\,-\left(i \Delta_2+\frac{\gamma_2}{2}+\frac{\gamma_1}{2}\right) \rho_{r_2 r_1}, \\
\dot{\rho}_{r_2 g} \  = &\ i \frac{\Omega_m}{2} \rho_{r_1 g}-i \frac{\Omega}{2} \rho_{r_2 r_1}\notag\\&\,-\left(i\left(\Delta_1+\Delta_2\right)+\frac{\gamma_2}{2}\right) \rho_{r_2 g}, \\
\dot{\rho}_{r_1 g} \  = &\ i \frac{\Omega}{2}\left(\rho_{g g}-\rho_{r_1 r_1}\right)+i \frac{\Omega_m}{2} \rho_{r_2 g}\notag\\&\,-\left(i \Delta_1+\frac{\gamma_1}{2}\right) \rho_{r_1 g}.
\end{align}
In the mean-field treatment, we solve the Lindblad master equation by replacing the detuning $\Delta_1$ with \(\Delta_1 \rightarrow \Delta_1 + \Delta_{\text{shift}} - V\rho_{r_1r_1}\). By changing the microwave power, we can obtain the series lines of Rydberg population $\rho_{r_1r_1}$. 

We consider the loss of the microwave field and use the substitution relationship between microwave power \( P \) and Rabi frequency \( \Omega_m \), given by the formula \( \Omega_m = \frac{\mu}{\hbar\sqrt{\varepsilon_0 c A}} 10^{\frac{P}{20}-1.5} \) in the theoretical part, where \(A\) represents the square of the region where Rydberg atoms interact with the microwave. This allows us to obtain the dashed lines in Fig.~\ref{Fig2}(b) for the phase diagram of the system. The relationship between output microwave power on the signal generator settings and the actual microwave power at the atoms has been measured, presenting a loss of 12.91 dBm from the signal generator to the atoms. By applying both forward and backward scans on \( \Delta_1 \), we can generate a hysteresis loop represented by the blue and red lines. In this process, we select the closest steady state at \( \rho_{r_1r_1}[t + \Delta t] \) as the next step in the evolution of \( \rho_{r_1r_1}[t] \). Consequently, this approach enables us to construct the hysteresis loop effectively.

\end{document}